\documentclass[conference]{IEEEtran}
\IEEEoverridecommandlockouts
\usepackage{cite}
\usepackage{amsmath,amssymb,amsfonts}
\usepackage{algorithmic}
\usepackage{booktabs}
\usepackage{graphicx}
\usepackage{textcomp}
\usepackage{xcolor}
\usepackage{geometry}
\begin{document}

\title{ExoSpikeNet: A Light Curve Analysis Based Spiking Neural Network for Exoplanet Detection\\
}

\author{\IEEEauthorblockN {Maneet Chatterjee$^{1}$, Anuvab Sen$^{2}$ and Subhabrata Roy$^{3}$}
\IEEEauthorblockA{$^{1}$ Department of Mechanical Engineering, IIEST Shibpur, Howrah - 711103, India\\$^{2,3}$ Department of Electronics and Telecommunication Engineering, IIEST Shibpur, Howrah - 711103, India\\
Email: maneet2018@gmail.com $^{1}$, sen.anuvab@gmail.com $^{2}$ and subhabrata\_ece@yahoo.com $^{3}$}\vspace{-1 cm}}

\maketitle

\begin{abstract}
Exoplanets are celestial bodies orbiting stars beyond our Solar System. Although historically they posed detection challenges, Kepler's data has revolutionized our understanding. By analyzing flux values from the Kepler (K2) Mission, we investigate the intricate patterns in starlight that may indicate the presence of exoplanets. This study has investigated a novel approach for exoplanet classification using spiking Neural Networks (SNNs) applied to the data obtained from the NASA Kepler (K2) mission. SNNs offer a unique advantage by mimicking the spiking behavior of neurons in the brain, allowing for more nuanced and biologically inspired processing of temporal data. Experimental results showcase the efficacy of the proposed SNN architecture, excelling in terms of various performance metrics such as accuracy, F1 score, precision, and recall.
\end{abstract}


\section{INTRODUCTION}

The rise of life on Earth has sparked an enduring quest to explore existence, both on our planet and across the vast cosmos. Technological advancements have played a pivotal role in accelerating progress in this age-old pursuit. Notably, the integration of machine learning, deep learning, and advanced analytical techniques has revolutionized scientific inquiry \cite{sen2023comparative} \cite{mazumder2024benchmarking} \cite{Sen_2023}. In the realm of astronomical and space science research, the past two decades have witnessed remarkable synergy among these technologies, facilitating streamlined matching, alignment, and analysis of vast datasets to unveil compelling evidence suggesting that we might not be the sole living beings in the universe. This paper aims to harness the power of robust and efficient machine and deep learning algorithms to determine whether a given star hosts an exoplanet within its orbit. While conventional machine learning and deep learning models have been instrumental in advancing the field, our approach transcends established norms. In addition to implementing classical deep learning models such as Convolutional Neural Networks or CNNs and Visual Geometry Groups or VGGs, we introduce our novel model known as the Spiking Neural Network (SNN) \cite{ghosh2009spiking}. This new application offers unique advantages in the field of exoplanetary data analysis that are expected to aid in future astronomical data analysis. Unlike other traditional models, which rely solely on static data inputs, Spiking Neural Networks will skillfully capture dynamic patterns and observe temporal dependencies that are present in large astronomical datasets. This acute sensitivity of the model aligns with the characteristics of the photon-flux data, where minute changes in photon intensity over time indicate probable exoplanetary transits.

The introduction of the Spiking Neural Network contributes to the addition of a new architecture in astronomical data analysis, that brings forth advantages that include increased adaptability to complex patterns, improved accuracy in detecting transient signals, and enhanced interpretation of features within the photon-flux data. By incorporating this model into our research methodology, we aim to not only improve the accuracy of exoplanet detection in the current analysis method but also pave the way for more robust and versatile approaches in future studies. While other machine learning and deep learning models have laid the foundation for accurate exoplanet detection, the integration of the spiking Neural Network into our model introduces a newer approach. Through its unique temporal processing capabilities, this model is poised to unlock new dimensions in the analysis of astronomical data, providing a promising avenue for future advancements in our understanding of exoplanetary systems. With regards to this, we aim to address two main queries:

\begin{itemize}
    \item What methods can we utilize to deduce the existence of an exoplanet solely based on the flux data emitted by a star?
    
\end{itemize}
\begin{itemize}
    \item In what way will machine learning and deep learning impact the precise forecasting of exoplanets using the dataset?
\end{itemize}

To the best of our knowledge, the use of Spiking Neural Networks (SNNs) for exoplanet detection seems new and no previous literature investigates the utilization of energy-efficient deep learning for the given task. The results are compared with other state-of-the-art deep learning and machine learning models.


\section{PRELIMINARIES}

In this section, we explore various machine learning (ML) and deep learning (DL) architectures, providing an overview of various models for exoplanet detection. Specifically, we focus on Gradient Boosting Machines (GBM), Random Forest Classifiers, Gaussian Naive Bayes, 1-Dimensional Convolutional Neural Networks (1D CNN), 2-Dimensional Convolutional Neural Networks (2D CNN), Multilayer Perceptrons (MLP) and proposed one \cite{10371846} \cite{sendifferential}.

\subsection{Machine Learning Algorithms} 

\subsubsection{Naïve Bayes}
The Naïve Bayes algorithm is a set of supervised learning algorithms that is based on Bayes' Theorem \cite{chandra2023detection}. We have implemented the Gaussian Naïve Bayes algorithm, which was imported from the scikit-learn 1.4.0 library. In this algorithm training and classification of data are done in accordance with the multivariate Gaussian distributions, where although multiple features may be present, they are considered to be binary-valued Gaussian booleans. Based on the binarized parameter, the algorithm may binarize the input data when needed. The formula describing the Gaussian Naïve Bayes algorithm is:

  $ P(x_i|y) = P(x_i|y)x_i+(1-P(x_i|y))(1-x_i) $

\vspace{0.2cm}
\subsubsection{Gradient Boosting Machines (GBM)}
Gradient Boosting algorithms have been widely accepted for classification tasks in exoplanet detection \cite{Malik_2021}. They employ an ensemble learning approach that fuses the predictions of numerous weak learners, usually decision trees and then construct a robust prediction model. The formula for predicting the target variable \( \hat{y} \) can be articulated as follows:

\[ \hat{y}_i = \sum_{k=1}^K f_k(x_i), \]

where: \( \hat{y}_i \) is the predicted value for the \( i \)-th instance, \( K \) is the number of trees in the ensemble, \( f_k(x_i) \) is the prediction of the \( k \)-th tree for the \( i \)-th instance \( x_i \). In practice, the prediction of each tree is weighted by a learning rate \( \nu \) and added to the predictions of previous trees. The prediction formula for a GBM with a learning rate \( \nu \) and shrinkage regularization is:
\[ \hat{y}_i = \sum_{k=1}^K \nu \cdot f_k(x_i), \]

where: \( \nu \) is a hyperparameter typically set to a value between 0 and 1.
 
\subsubsection{Random Forest}
Random Forest is an ensemble machine learning method that adds the predictions of multiple decision trees to arrive at a final prediction score. This approach typically yields more robust and accurate predictions compared to individual trees. 
The general formula for making predictions can be described as follows:

   \[ \hat{y} = \text{mode}\left( f_1(x), f_2(x), ..., f_n(x) \right) \]
   where: \(\hat{y}\) is the prediction class label, \(f_i(x)\) represents the prediction of the \(i^{th}\) decision tree in the forest considered as input \(x\), and \(\text{mode}\) refers to the most frequent class label among the predictions of all trees.
   \[ \hat{y} = \frac{1}{n} \sum_{i=1}^{n} f_i(x) \]
   where: \(\hat{y}\) is the predicted output (mean prediction), \(f_i(x)\) represents the prediction of the \(i^{th}\) decision tree in the forest for input \(x\), and \(n\) is the total number of trees in the forest.
   
This machine learning model will aid in exoplanet classification by integrating an ensemble of decision trees to accurately decipher patterns in the Kepler flux data. Therefore enhancing the identification process of exoplanetary signatures amidst stellar flux variations \cite{McCauliff_2015}.

\subsection{Deep Learning Models}

\subsubsection{1-Dimensional Convolutional Neural Networks}
1-dimensional Convolutional Neural Networks (1-D CNNs) are classical deep learning models that are commonly employed for analyzing sequence data, notably time-series data like light curves in exoplanet detection \cite{koning2019reducing}. In exoplanet research, 1-D CNNs have been employed to automatically detect transit signals in light curves, hence capitalizing on their capacity to learn hierarchical representations of transient features. The formula for computing the output of a 1-D CNN is given as follows:

\[ z_i = f(\sum_{j=1}^m (x_{i+j} \ast w_j) + b), \]

where : \( z_i \) is the output of the \( i \)-th neuron, \( x_{i+j} \) are the input values in the receptive field, \( w_j \) are the filter weights, \( b \) is the bias term, \( f(\cdot) \) is our activation function, and \( \ast \) denotes the convolution operation. 

Thus, 1-D CNN provides a robust tool for automatic feature extraction and classification in exoplanet detection.

\subsubsection{2-Dimensional Convolutional Neural Networks}

2-dimensional Convolutional Neural Networks (2-D CNN) is an exceptional tool for image-related tasks, as they extract spatial features through multiple convolutional layers. By using its capacity to capture spatial relationships and hierarchical patterns within the input data, such as lightcurves or periodograms, a 2-D CNN can aid in the classification of exoplanet data from the Kepler flux dataset \cite{Chintarungruangchai_2023}. The formula for computing the output of a 2-D CNN model is given as follows:

\[ Z_{ij}^{(l)} = f\left(\sum_{m=1}^{M} \sum_{n=1}^{N} \sum_{c=1}^{C} W_{mnck}^{(l)} \cdot X_{(i+m)(j+n)c}^{(l-1)} + b_k^{(l)}\right) \]

Where: \( Z_{ij}^{(l)} \) is the activated neuron \( (i, j) \) in the \( l \)- convolutional layer, \( f(\cdot) \) is the activation function (e.g., ReLU),  \( W_{mnck}^{(l)} \) is the weight parameter for the \( (m, n) \)-th filter in the \( k \)-th channel of the \( l \)-th layer, \( X_{(i+m)(j+n)c}^{(l-1)} \) is the input from the \( (l-1) \)-th layer, \( b_k^{(l)} \) is the bias term for the \( k \)-th filter in the \( l \)-th layer, and \( M \) and \( N \) are the spatial dimensions of the filter with \( C \) being the number of channels in the input image.

\subsubsection{Multilayer Perceptron (MLP)}
A Multilayer Perceptron (MLP) \cite{Johnsen_2020} comprises a Feed-Forward Neural Network model, characterized by the presence of many neuron layers. 
The formula for computing the output of a single neuron in an MLP is as follows:

\[ z_j = \sum_{i=1}^n w_{ij} \cdot x_i + b_j, \]
where: \( z_j \) is fed to the activation function of neuron \( j \) in the current layer, \( w_{ij} \) is the weight of the connection between the neurons \( i \) in the previous layer and neurons \( j \) in the current layer, \( x_i \) is the output of the neuron \( i \) in the previous layer, \( b_j \) is the bias term associated with the neuron \( j \), \( n \) is the number of neurons in the previous layer.


\subsection{Spiking Neural Networks (SNNs)}

The Spiking Neural Network (SNN) is a type of neural network model that closely represents the behavior of biological neurons and their communication network through the generation of discrete, asynchronous spikes or action potentials as said in \cite{pritchard2023rfi}. Unlike classical neural network models where information is processed continuously, SNN architecture operates on discrete timesteps, with neurons firing spikes when their membrane potential reaches a certain threshold value. Fig. 1 illustrates the basic structure of a spiking Neural Network, consisting of input, hidden, and output layers which are interconnected by synapses. Artificial neurons in the network receive input signals, generate spikes, and transmit information through synaptic connections to produce output responses. A very basic formula for SNN using a leaky integrate-and-fire (LIF) model is given below:

\[
\tau_m \frac{dV}{dt} = -(V(t) - V_{rest}) + R I(t)
\]

Where: \( V(t) \) is the membrane potential of the neuron at time \( t \), \( V_{rest} \) is the resting membrane potential, \( \tau_m \) is the membrane time constant, \( R \) is the membrane resistance, and \( I(t) \) is the input current to the neuron.

When the membrane potential reaches a certain threshold value \( V_{th} \), the neuron emits a spike and its membrane potential is then reset to a resting state. Spiking neural Networks offer several advantages for processing temporal data such as lightcurve data from astronomical observations like the Kepler mission because:

\begin{itemize}
    \item Temporal processing: An SNN model is inherently suited to capture temporal patterns and trends in data, making it well-suited for processing lightcurve data from Kepler observations, where temporal dependencies play a crucial role in detecting exoplanetary transits.
\end{itemize}

\begin{itemize}
    \item Robustness to Noisy data: The event-based nature of spiking Neural Networks allows them to filter out noise and extract relevant features from the given data, potentially improving their robustness to noise and disturbances present in astronomical observations.
\end{itemize}

\begin{figure}[h!]
    \centering    \includegraphics[width=0.95\linewidth, height = 3.4 cm]{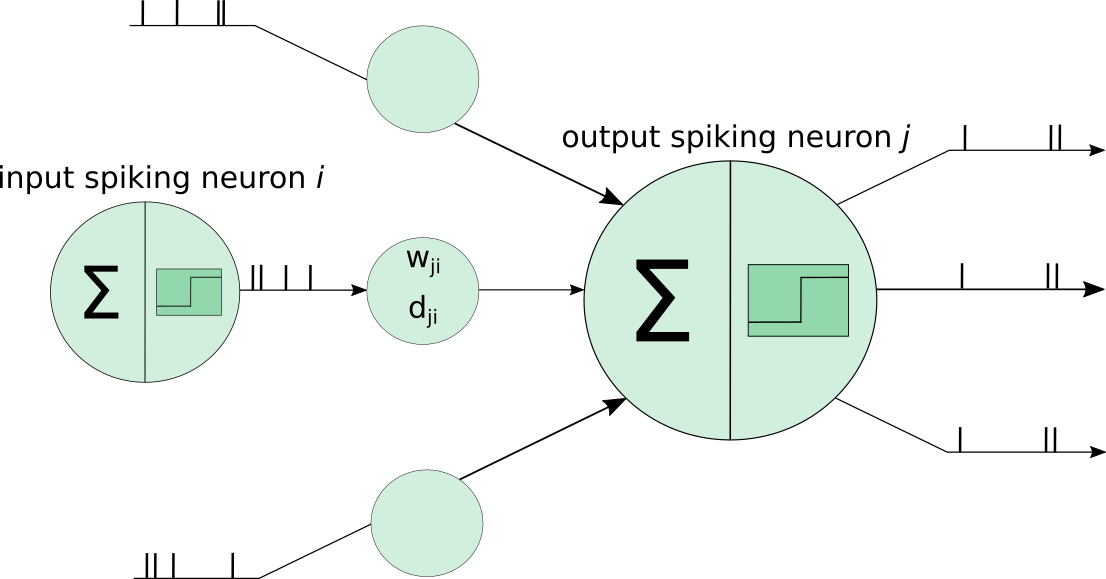}
    \caption{Spiking Neural Network Architecture}
    \label{fig: }
\end{figure}


\section{PROPOSED METHODOLOGY}

In this section, we provide a detailed explanation of our novel spiking Neural Network model devised for the prediction and classification of exoplanets from the Kepler photon-flux dataset. Our proposed architecture comprises the SNN model, which is implemented on our pre-processed Kepler flux dataset. The dataset comprises numerous entries, each corresponding to a multitude of intensity measurements captured from a star. A categorical label accompanies each set of measurements, denoting the presence or absence of an exoplanet. The labels are encoded as binary values: 2 indicating the presence of an exoplanet and 1 representing its absence. Although primarily aesthetic, we simplify the encoding interpretation (1 for presence, 0 for absence). Illustrated in Fig. 2 are two distinct light curves, randomly selected from the dataset, portraying stars with and without exoplanets, respectively.
\begin{figure}[h!]
    \centering
    \includegraphics[width=1\linewidth]{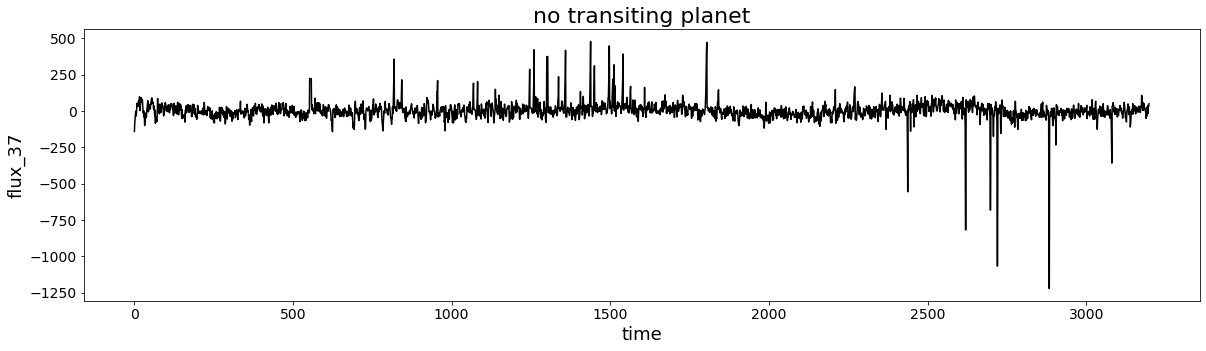}
\end{figure}
\begin{figure}[h!]
    \centering
    \includegraphics[width=1\linewidth]{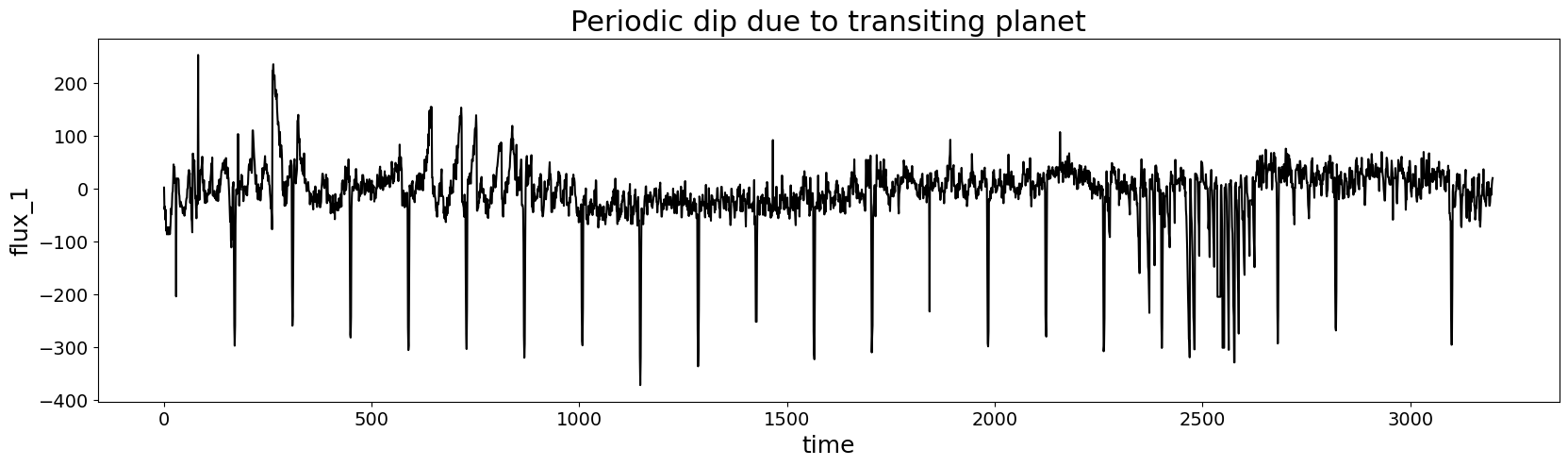}
    \caption{(a) Lightcurve of Star without exoplanet (b) Lightcurve of Star with an exoplanet.}
    \label{fig:enter-label}
\end{figure}
We start by conducting preprocessing on the dataset, followed by the generation of two random plots depicting stars with and without exoplanets. Scaling techniques were implemented to ensure uniformity in light curve intensity values, while outlier handling was utilized to identify and mitigate anomalies. Additionally, Gaussian filtering was applied to smooth the curve and enhance underlying patterns. These preprocessing steps are instrumental in enhancing data quality for accurate analysis of astronomical phenomena \cite{Davoudi_2020}.

The most interesting feature observed in light curve (Fig. 2b) is the periodic dip in brightness of the star, indicative of the transit of an exoplanet across the face of the star as observed from the Earth. These dips occur at regular intervals and are a characteristic feature of the orbital period of the exoplanet, which can be found in subsequent analysis. On the other hand, the absence of any periodic dip in Fig. 2a and the nearly constant flux change with respect to time suggests the absence of an exoplanet. A few random dips in flux value towards the end are likely attributed to instrumental effects within the atmosphere or telescope.

The periodogram analysis of Fig. 3b with an exoplanet reveals a dominant frequency of 0.0144, corresponding to a period of approximately 69.4 days. The high power-to-median power ratio of 479.36 suggests a significant signal amidst the noise, indicating a strong periodic pattern likely associated with the presence of an exoplanet in the observed data. For Fig. 3a, the absence of an exoplanet in the periodogram suggests that the detected peak at a frequency of 0.0003 corresponds to a periodic signal inherent in the data. With a corresponding period of approximately 3333.3 days, this signal likely represents a recurring astronomical phenomenon or instrumental artifact. The high ratio of the maximum power to the median power (51.87) indicates a significant peak relative to the background noise, reinforcing the absence of an exoplanet.

\begin{figure}[h!]
    \centering
    \includegraphics[width=1\linewidth]{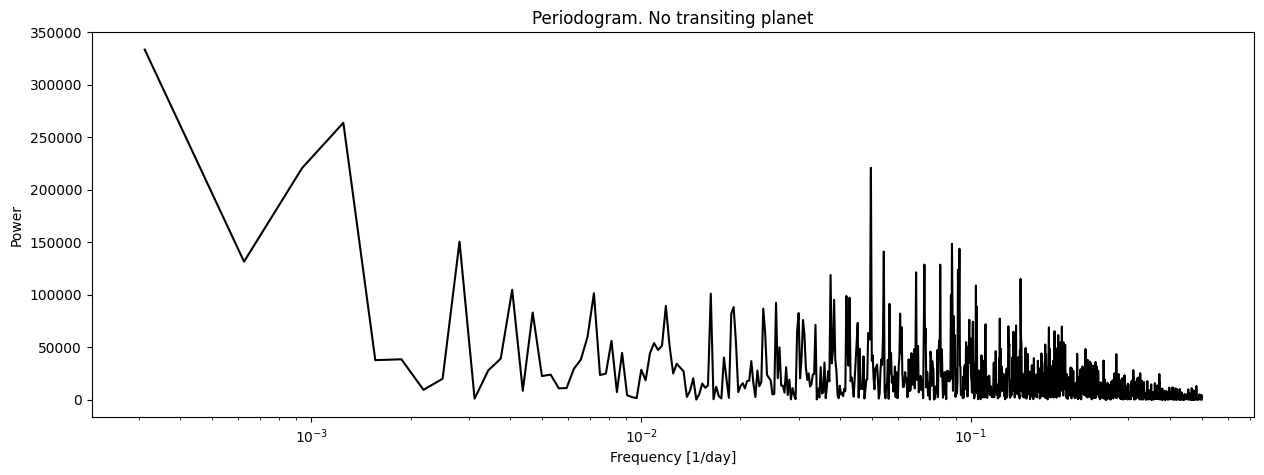}
\end{figure}

\begin{figure}[h!]
    \centering
    \includegraphics[width=1\linewidth]{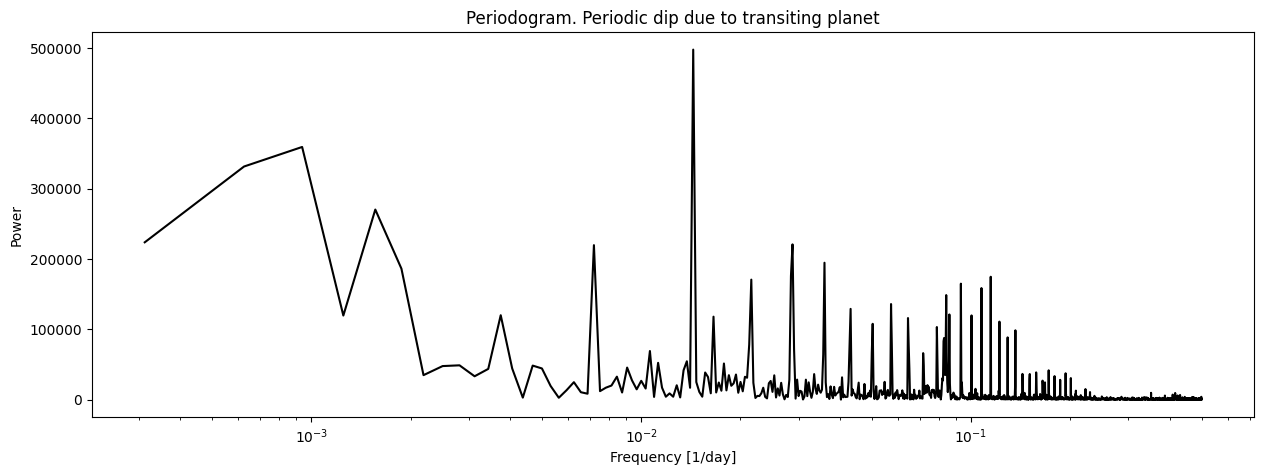}
    \caption{(a) Periodogram of the Star without exoplanet. (b) Periodogram of the Star with an exoplanet.}
    \label{fig:enter-label}
\end{figure}

To implement the spiking neural network, We commence by transforming the pre-existing Convolutional Neural Network model into a spiking Neural Network (SNN) utilizing Nengo-DL, a library tailored for constructing and simulating SNNs. This conversion procedure entails transferring the CNN architecture and weights onto a spiking neuron network framework. Subsequently, we preprocess the dataset to be fed into the SNN, ensuring its structure aligns appropriately with timesteps. This requires reshaping the input data to incorporate timesteps, a prerequisite for temporal processing within the SNN. Both the training and testing datasets should undergo corresponding transformations.
\begin{figure}[h!]
    \centering
    \includegraphics[width=1\linewidth]{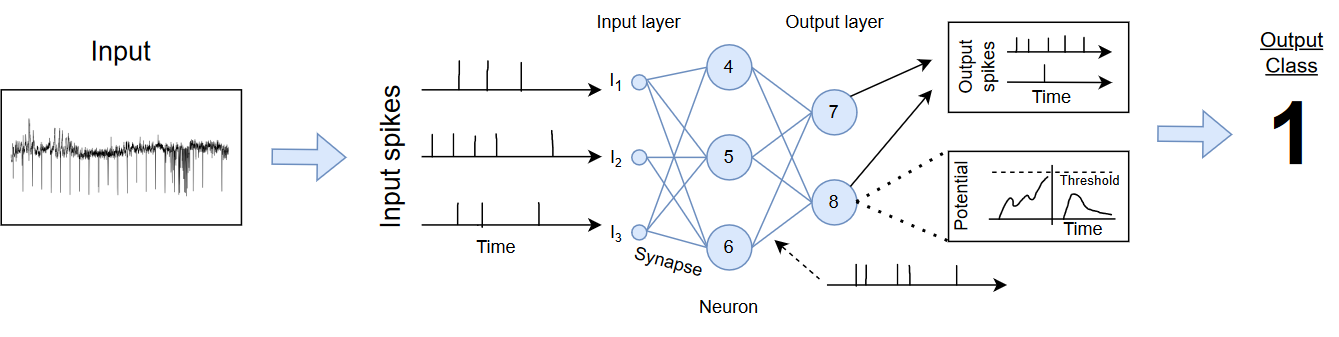}
    \caption{Architecture for evaluating spiking Neural Networks}
    \label{fig:enter-label}
\end{figure}
The outcomes derived from training the spiking Neural Network (SNN) using Nengo-DL are explained in Fig. 4, which demonstrates the architecture and methodology employed. The figure outlines the architecture employed for assessing spiking Neural Networks, featuring input layers, hidden layers, and output layers interconnected through synapses. Neurons within these layers process input signals, generating spike responses that propagate through the network, ultimately yielding output prediction. The input that is fed to the network is the transformed lightcurve of stars, which upon processing and prediction will generate 1 or 0 as considered earlier.
In this context, the successful training of the SNN resonates with the network architecture outlined in the caption. The SNN, characterized by its spiking neuron-based computation, can be likened to the convolutional layers with rank order coding as detailed in the caption. Furthermore, the training process of the SNN employs optimizing network parameters to adeptly capture features and patterns from input data, similar to the processing steps mentioned in the caption.

\section{EXPERIMENTAL SETUP}

\subsection{Dataset Description}

The dataset utilized, in this paper, is publicly available and derived from observations conducted by NASA's Kepler mission. The Kepler Space Telescope primarily detected exoplanets by scrutinizing variations or abrupt fluctuations in the flux or luminosity levels of stellar systems. As a result, the dataset comprises flux measurements obtained from various stars over specific time intervals, including some that constitute multi-planet systems. Notably, the dataset employed herein is partitioned into two distinct segments:

\begin{itemize}
    \item Training set - It comprises 5087 stars, among which 37 are confirmed exoplanets, while 5050 stars are devoid of exoplanets. Each of the 5087 stars is associated with 3198 confirmed observations of light intensity or flux values over the designated time-period.
\end{itemize}

\begin{itemize}
    \item Testing set - It is composed of 570 stars, with the presence of 5 confirmed exoplanets. All 570 stars are characterized by 3198 confirmed light intensity observations over the stipulated time-period.
\end{itemize}


\subsection{Dataset Preprocessing}
The dataset underwent comprehensive pre-processing to enhance computational efficiency and ensure consistency in feature scales. Utilizing various scaling techniques such as Standardization, Normalize, MinMax Scaler, and MaxAbs Scaler allowed for diverse approaches to feature scaling tailored to specific requirements. In addressing the dataset's high dimensionality (3198 features), Feature Engineering and Principal Component Analysis (PCA) were employed for dimension reduction post-scaling, optimizing computational resources while retaining pertinent information. Furthermore, to mitigate noise and filtering, the Fast Fourier Transform (FFT) was applied, followed by signal smoothing with the aid of Savitzky-Golay filter (Savgol). Subsequent normalization of the signal, along with the utilization of a robust scaler, effectively handled outliers. This meticulous preprocessing pipeline aims to optimize the dataset for subsequent analysis and model development, enhancing computational efficiency and ensuring robust performance. Additionally, focusing on the first 5 columns, which account for almost 75\% of the data, we applied streamlined model training and testing while preserving critical information. The dataset has been partitioned into training (70\%), testing (20\%), and validation (10\%) subsets.

\section{RESULTS AND DISCUSSION}
In this section, we explore the exciting possibilities unlocked by leveraging spiking Neural Network (SNN) models to improve classification accuracy and reliability.
\begin{figure}[h!]
    \centering
    \includegraphics[width=1\linewidth]{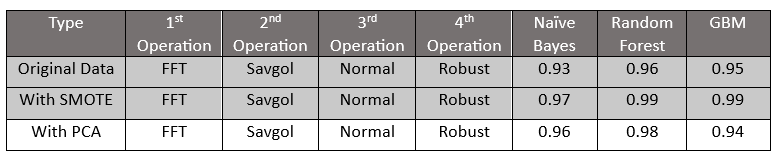}
    \caption{Table showing F1 scores of testing data with different combinations of Scalers and Normalizations with machine learning models}
    \label{fig:enter-label}
\end{figure}

We analyze the results obtained from various machine and deep learning techniques and contrast them with the outcomes achieved using SNN on our dataset. Throughout our investigation, we introduce our unique approach with SNN and thoroughly compare it with established models like 1D and 2D CNN, MLP, and VGG-16.

For machine learning models, devoid of feature engineering and relying solely on the original testing dataset, all models exhibited suboptimal performance. Hyperparameter tuning on the original data, as depicted in Fig. 5, yields unsatisfactory results. However, upon incorporating feature engineering with the original dataset, significant enhancements in model performance were observed. Notably, employing SMOTE results in better classification scores across almost all models. Detailed results obtained through feature engineering are illustrated in Fig 5.
\begin{table}[h!]
  \centering
  \caption{Machine Learning Model Comparison}
  \label{tab:model_performance}
  \begin{tabular}{lcccc}
    \toprule
    \textbf{Model} & \textbf{ Accuracy} & \textbf{ Precision} & \textbf{ F1 Score} & \textbf{ Recall} \\
    \midrule
    Naïve Bayes    & 0.98 & 0.94 & 0.97 & 0.96 \\
    RandomForest   & 0.98 & 0.97 & 0.97 & 0.98 \\
    GBM            & 0.97 & 0.98 & 0.98 & 0.96 \\
    \bottomrule
  \end{tabular}
\end{table}

Performance parameters, obtained from the specified machine-learning model, are outlined in Table I, alongside the innovative application of feature engineering. By integrating deep learning models into our training, testing and validation data, the results have significantly improved in terms of accuracy. The proposed model is trained with the help of hyperparameter tuning.
\begin{table}[h!]
    \centering
    \caption{Deep Learning Model Comparison}
    \begin{tabular}{@{}lllll@{}}
        \toprule
        Model & Accuracy & Precision & F1 Score & Recall \\ \midrule
        1D CNN & 0.97 & 0.96 & 0.97 & 0.97 \\
        2D CNN & 0.98 & 0.97 & 0.98 & 0.98 \\
        VGG 16 & 0.94 & 0.97 & 0.96 & 0.95 \\
        MLP & 0.95 & 0.96 & 0.95 & 0.95 \\
        SNN & 0.99 & 0.98 & 0.98 & 0.99 \\ \bottomrule
    \end{tabular}
\end{table}

A comparison in between different deep learning models is tabulated in Table II. From Table I and Table II, it can be observed that proposed spiking neural network model outperforms all other state-of-the-art models with an accuracy of 99\% while maintaining precision, recall and F1 score at the desired level.
\begin{figure}[h!]
    \centering    \includegraphics[width=0.85\linewidth, height = 4.8 cm]{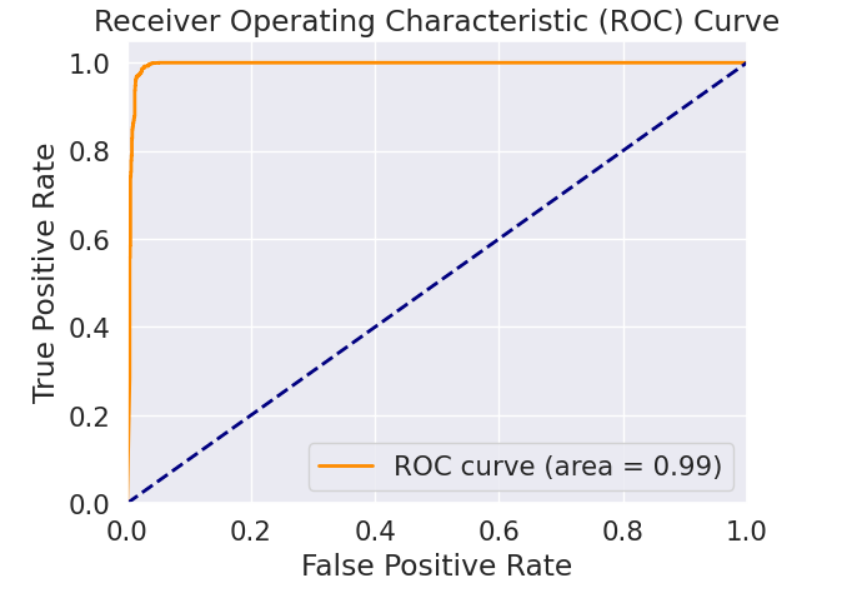}
    \caption{Area under ROC curve for SNN}
    \label{fig:enter-label}
\end{figure}

Furthermore, we explore the performance of the SNN model over SMOTE data. Receiver Operating Characteristics (ROC) curve has been displayed in Fig. 6 with an Area Under the Curve (AUC) equivalent to 99\%. As depicted, the curve illustrates the model's performance across various classification thresholds, confirming its efficiency in distinguishing stars with and without exoplanets.
\begin{figure}[h!]
    \centering
    \includegraphics[width=1\linewidth, height= 5cm]{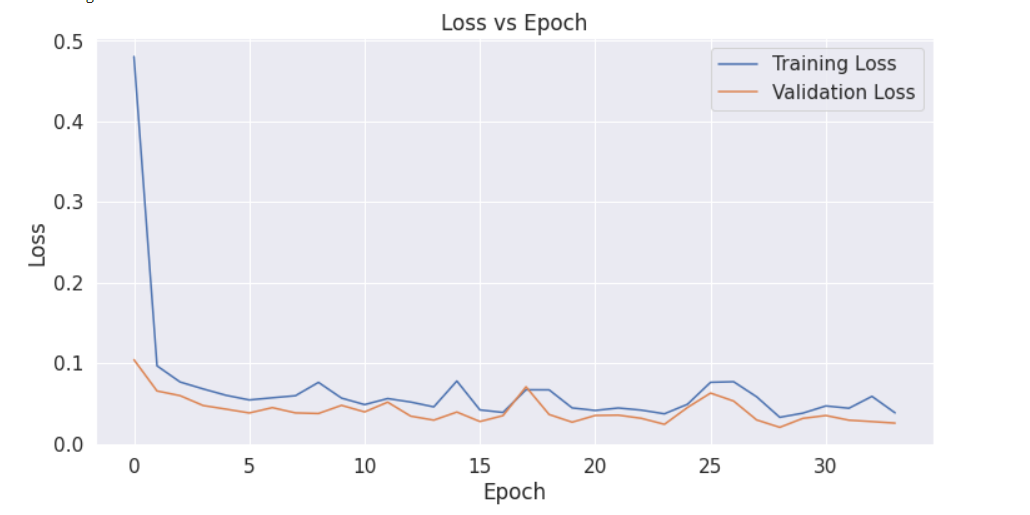}
    \caption{Training loss vs epochs plot}
    \label{fig:enter-label}
\end{figure}

The training loss vs. epochs curve has been presented in Fig. 7 to assess the performance of the spiking Neural Network in exoplanet detection and classification. The confusion matrix of the spiking neural network is shown in Fig. 8. This graphical representation sheds light on the precise evaluation of classification accuracy and helps us to identify the misclassifications; hence facilitating an optimized model formation with advanced decision-making skills, applicable for real-world problems.
\begin{figure}[h!]
    \centering
    \includegraphics[width=1\linewidth, height = 6 cm]{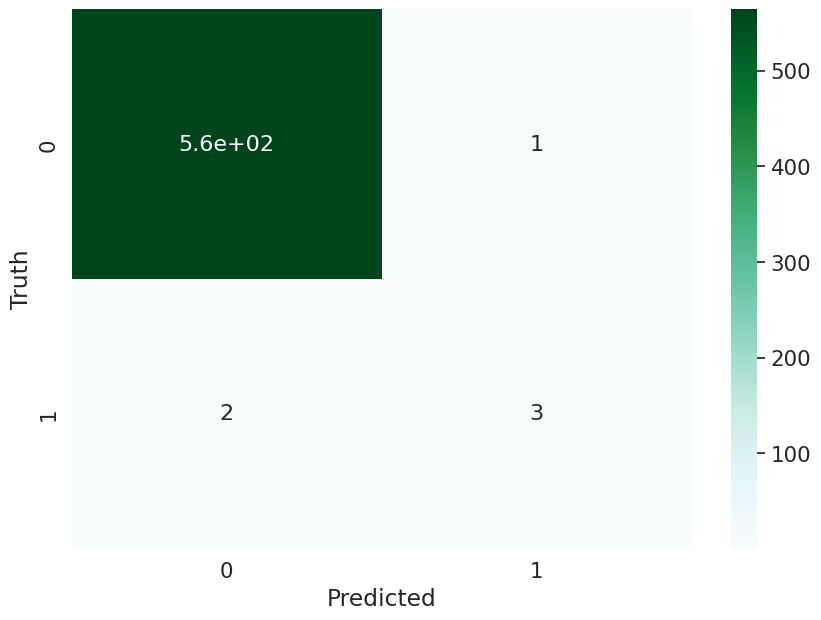}
    \caption{Confusion Matrix of the SNN architecture}
    \label{fig:enter-label}
\end{figure}



\section{CONCLUSION AND FUTURE WORK}


This paper introduces a novel Spiking Neural Network (SNN) architecture for the given exoplanet detection task. The model performance has been compared to other classical deep learning and machine learning models developed for the classification of exoplanets from the Kepler mission dataset. The superiority of SNN with an exceptional test accuracy of approximately 99.52\% plausibly lies in its ability to emulate the spiking behavior of biological neurons, enabling more nuanced processing of temporal data and capturing complex patterns inherent in astronomical datasets.

Future research may be directed toward the prospect of combining Spiking Neural Networks (SNNs) with Recurrent Neural Networks (RNNs) which may present a promising avenue for future research. By leveraging the temporal processing capabilities of RNN, this hybrid model can effectively
capture temporal dependencies and subtle variations in data, making it well-suited for gaining additional insights into time-varying analyses on varying flux data.

\bibliographystyle{IEEEbib}
\bibliography{strings,refs}

\end{document}